# Spectroscopic evidence of symmetry breaking in the superconducting vortices of $UTe_2$

Zhongzheng Yang[1,*], Fanbang Zheng[1,*], Dingsong Wu[2,*], Bin-Bin Zhang[3,*], Ning Li[3], Wenhui Li[1,4], Chaofan Zhang[3,†], Guang-Ming Zhang[1], Xi Chen[5], Yulin Chen[1,2,4,†], Shichao Yan[1,4,†]

[1]*State Key Laboratory of Quantum Functional Materials, School of Physical Science and Technology, ShanghaiTech University, Shanghai, China*

[2]*Department of Physics, University of Oxford, Oxford, UK*

[3]*Nanhu Laser Laboratory, Changsha, China*

[4]*ShanghaiTech Laboratory for Topological Physics, ShanghaiTech University, Shanghai, China*

[5]*State Key Laboratory of Low-Dimensional Quantum Physics, Department of Physics, Tsinghua University, Beijing, China*

[*]*These authors contributed equally*

[†]*Email: yanshch@shanghaitech.edu.cn; yulin.chen@physics.ox.ac.uk; hjroland@163.com*

## ABSTRACT

**The recently discovered heavy-fermion superconductor, $UTe_2$, is an excellent candidate for spin-triplet superconductors where electrons form spin-triplet Cooper pairs with spin $S = 1$ and odd parity. Unconventional superconductivity often hosts unconventional vortices. Yet, the vortex core and lattice in $UTe_2$ have not been directly visualized and characterized. Here, by using ultralow-temperature scanning tunneling microscopy and spectroscopy, we study the superconducting vortices on the (0−11) surface termination of $UTe_2$ with an out-of-plane external magnetic field. At the center of the vortex core, we observe a robust zero-energy vortex-core state which exhibits a cigar-shaped spatial distribution and extends to ~30 nm along the [100] direction (crystallographic *a* axis) of $UTe_2$. Along the direction perpendicular to [100], the superconducting gap is deeper and the coherence peak on one side of the vortex core is stronger than on the opposite side, and they are even enhanced in comparison with those under zero field. Due to the anisotropy of magnetic susceptibility in $UTe_2$, the asymmetric d*I*/d*V* spectra on the two sides of the vortex core result from the interplay between the magnetization-induced bound current and supercurrent around the vortex core. Our work reveals the important role of magnetization in the vortex behaviors of $UTe_2$ and provides essential microscopic information for understanding its superconducting properties in magnetic field.**

## INTRODUCTION

Spin-triplet pairing is a fascinating phenomenon, which has been predicted to exhibit many novel electronic properties, including the fractionalized electronic states and topological edge modes [1-6]. Because of the coexistence of magnetism and superconductivity, the U-based heavy-fermion superconductors are particularly promising for the realization of the spin-triplet pairing [7-10]. In this context, $UTe_2$ has shown strong evidence as a spin-triplet superconductor [11-20]. Although the identity of the superconducting order parameter in $UTe_2$ is still under debate, several unusual superconducting properties have been reported in $UTe_2$, including a large and highly anisotropic upper critical field that exceeds the Pauli limit [11-13], multiple superconducting regimes under extreme magnetic fields [13], a negligible change in the temperature-dependent nuclear magnetic resonance (NMR) shift cooling through superconducting transition temperature (although the recent NMR measurements show large reduction in the *a*-axis Knight shift for higher-quality $UTe_2$) [15, 21], coexistence of superconductivity and ferromagnetic fluctuations from muon-spin relaxation measurements [16], chiral in-gap states at step edges in the low-temperature scanning tunneling microscopy and spectroscopy (STM/STS) measurements [17]. All these observations provide strong evidence in support of spin-triplet superconductivity in $UTe_2$.

Superconductivity in $UTe_2$ emerges upon cooling from a paramagnetic state and coexists with strong ferromagnetic fluctuations [11, 16, 22]. As a type-II superconductor, when a magnetic field (larger than the lower critical field but lower than the upper critical field) is applied to $UTe_2$, the magnetic field penetrates into $UTe_2$ in the form of vortices which consist of both magnetic fluxes and circulating supercurrents [23]. The ferromagnetic fluctuations in the vortex of $UTe_2$ can be influenced by the magnetic field within the vortex. This would result in unique vortex properties which are absent for the conventional superconducting vortices. As mentioned above, $UTe_2$ indeed shows several unusual behaviors in magnetic field [11-13], and their origins still remain mysterious. Directly probing the vortices in $UTe_2$ is an important step for understanding the unconventional superconducting properties of $UTe_2$ in magnetic field. The vortex core and lattice in superconductors can be directly probed by low-temperature and high-magnetic-field STM/STS technique [24-29]. Despite its low superconducting transition temperature and large residual density of states near zero energy [17, 30-32], the direct observation of the vortex core and lattice in $UTe_2$ still remains elusive.

**RESULTS**

Here we report an ultralow-temperature STM study of vortex lattice and vortex-core states on the (0−11) surface of $UTe_2$ single crystals. Bulk single crystals of $UTe_2$ have orthorhombic crystal structure, and the superconducting transition temperature ($T_{sc}$) is about 2 K (Fig. S1 in Supplemental Material). $UTe_2$ single crystals typically cleave to show the (0−11) surface [17, 30-33]. Similar to the previous STM measurements, the typical STM topographies on the (0−11) surface [Figs. 1(b) and 1(c)] exhibit a chain-like structure where two rows of Te atoms orient along the [100] direction (crystallographic *a* axis) [17]. Differential tunneling conductance (d*I*/d*V*) probes the local density of states and can measure the superconducting gap near the Fermi level. In the d*I*/d*V* spectrum taken in energy range of ± 1 meV, we observe the superconducting gap with symmetric coherence peaks located around ± 0.25 meV. The superconducting gap is gradually suppressed as the temperature increases to $T_{sc}$ ~ 2 K [Fig. 1(e)].

Figures 1(g) and 1(i) show the linecuts of d*I*/d*V* spectra taken in different energy ranges and along the yellow arrow in Fig. 1(d). As shown in Fig. 1(g), in addition to the superconducting gap at the Fermi level, there is a peak-like feature at ~−4.5 mV, which may be related to the flat band derived from the *f* electrons in $UTe_2$. Although the d*I*/d*V* signal above the superconducting gap is more or less spatially uniform, the depth of the superconducting gap shows spatial dependence [Figs. 1(f) and 1(h)]. As shown in the d*I*/d*V* linecut profile with ± 1 meV energy range [Fig. 1(i)], the depth of the superconducting gap exhibits periodic spatial modulation, and the depth of the superconducting gap at the $Te_2$ chain is slightly larger than that at the $Te_1$ chain [Fig. 1(h)]. Although the size of the superconducting gap is similar to that reported in the previous STS measurements [17, 31], the depth of the superconducting gap measured in this work is significantly larger (Fig. S2 in Supplemental Material), which could be due to the lower measurement temperature (~30 mK lattice temperature) and slightly higher $T_{sc}$ for our $UTe_2$ single crystal. We note that although some properties of $UTe_2$ may depend on the quality of samples [34-38], STM is a local probe technique and the previously reported chiral edge states and charge density wave on the (0−11) surface of $UTe_2$ can be repeated in our STM measurements (Fig. S3 in Supplemental Material) [17, 30-33].

Having confirmed the zero-field superconductivity in $UTe_2$, we next investigate the vortex lattice and vortex-core states by performing the d$I$/d$V$ measurements with the external magnetic field perpendicular to the (0−11) surface. In this case, the magnetic field is along the direction with the angle offset ~24° from the crystallographic $b$ to $c$ axes [39]. The spatial distribution of the vortex core reflects the quasiparticle wave function and can be mapped out by performing the d$I$/d$V$ map measurements. Figure 2(a) shows the zero-energy d$I$/d$V$ map, and a cigar-shaped vortex core appears elongated along the [100] direction. Figure 2(b) is the d$I$/d$V$ map taken at the coherence peak energy (−0.25 mV) and in the same area as in Fig. 2(a). In the d$I$/d$V$ spectrum taken at the center of the vortex, we observe a zero-energy vortex-core state with the full width at half maximum ~0.2 mV [Fig. 2(d)]. Figures 2(e) and 2(f) are the d$I$/d$V$ linecut spectra taken along the [100] and [011] directions, respectively [denoted by the dashed arrows in Fig. 2(a)]. In the linecut of d$I$/d$V$ spectra along [011] direction, the zero-energy conductance peak is located within a narrow spatial range [Fig. 2(f)]. However, the zero-energy peak extends to ~30 nm along the [100] direction and does not split [Fig. 2(e)]. This indicates that the zero-energy vortex-core state is highly anisotropic, which can be explained by the anisotropy of the Ginzburg-Landau coherence length ξ along the two directions. The cigar-shaped vortex should be attributed to the anisotropic Fermi surface and the superconducting gap structure in $UTe_2$ [25]. By fitting the zero-energy conductance values as a function of position with an exponential decay, the extracted coherence lengths along the two directions are $\xi_1$ ~ 15 nm and $\xi_2$ ~ 5 nm, respectively (Fig. S4 in Supplemental Material).

Another prominent feature shown in the d$I$/d$V$ maps is that the d$I$/d$V$ signal on the right side of the vortex core (along the [011] direction from the vortex center) appears different from that on its left side [Figs. 2(a)-(c)]. The zero-energy d$I$/d$V$ signal on the right side of the vortex core is significantly lower than that on the left side [Figs. 2(a) and 2(c)], and the d$I$/d$V$ signal at the coherence peak energy (−0.25 mV) has higher intensity on the right side [Fig. 2(b)]. This indicates that the right side of the vortex shows deeper superconducting gap and stronger coherence peak, which can be clearly seen in the d$I$/d$V$ spectra taken on the two sides of the vortex core [Fig. 2(d)]. More surprisingly, the right-side superconducting gap and the coherence peak are even more prominent than those taken at zero magnetic field [Fig. 2(d)]. The enhancement of superconductivity appears within a ~20 × 20 nm$^2$ area on the right-side of the vortex core, and it induces inversion symmetry breaking along the [011] direction. To exclude the possibility that this inversion symmetry breaking near the vortex core is induced by local defects, we perform linecuts of d$I$/d$V$ spectra measurement without magnetic field along the same red-dashed arrow in Fig. 2(a), and no symmetry breaking feature is observed (Fig. S5 in Supplemental Material). We also note that the symmetry breaking feature only appears in the d$I$/d$V$ spectra within the superconducting gap energy range (Fig. S6 in Supplemental Material).

To reveal the evolution of the symmetry breaking near the vortex core with external magnetic fields, we perform the magnetic-field-dependent measurements for the vortices in $UTe_2$. Figures 3(a)-(e) show the zero-energy d$I$/d$V$ maps taken with different magnetic fields perpendicular to the (0−11) surface, and the density of the vortices is proportional to the strength of the magnetic field. Figure 3(f) shows the number of vortices as a function of external magnetic fields in a 100 × 100 nm$^2$ area, which indicates that each vortex carries one magnetic flux quanta ($\phi_0$ ~ 2.07 × $10^{-15}$ Wb). At low magnetic fields, such as 0.5 T, the symmetry breaking near each vortex core can be clearly seen [Fig. 3(a)]. As the magnetic fields increase to be above ~2 T, the vortices form the triangular lattice and the right side of a vortex often overlaps with the left side of the neighboring vortex. This makes the asymmetry near the vortex cores difficult to be distinguished [Figs. 3(b)-3(e)]. As the magnetic field increases up to ~2 T, the depth of the superconducting gap in the d*I*/d*V* spectra measured on the right side of the vortex gradually gets smaller [Fig. 3(k)]. At the same time, the intensity of the zero-energy vortex-core state decreases, which could be due to the vortex-vortex interaction [Fig. 3(j)]. Similar behavior has also been observed for the zero-energy vortex-core state in the iron-based superconductor (LiFeAs) [40].

During the d*I*/d*V* map measurements, we find that the vortices in $UTe_2$ are weakly pinned and easy to move. Figures 3(g)-3(i) are three typical zero-energy d*I*/d*V* maps taken in the same area and within 48 hours. We can see that although the vortices can be stable for a few hours to allow the d*I*/d*V* map measurements, the vortex lattice keeps changing for a longer time. The dashed ellipses in Figs. 3(g) and 3(h) mark the moved vortex cores captured in the d*I*/d*V* map measurements, and the vortex motion results in a break on a certain line of the d*I*/d*V* map. Interestingly, no matter where the vortices locate, the symmetry breaking is associated with all the vortex cores. This also rules out the possibility that the symmetry breaking is due to the local defects near the vortex core. The weakly pinning effect of the vortices in $UTe_2$ is consistent with the recent direct current resistivity measurements [41]. Furthermore, we find that this kind of symmetry breaking near the vortex core is independent of the direction of the out-of-plane magnetic field. Figures 4(a) and 4(b) are the zero-energy d*I*/d*V* maps taken on the same area with +1 T and −1 T magnetic fields, and they show the same asymmetry. This demonstrates that the symmetry breaking near the elongated vortex core is not due to the possible misalignment between the direction of magnetic field and the surface normal. Otherwise, when the out-of-plane direction of the magnetic field is reversed, the asymmetry near the vortex core should also reverse.

**DISCUSSION**

Finally, we discuss the possible origin of the symmetry breaking near the vortex core of $UTe_2$. First, the maximum magnetic field that we applied is about 5 T and perpendicular to the (0−11) surface. It is above the Pauli limit for $UTe_2$ (~3.7 T for $T_{sc}$ ~ 2 K), but it is below the strength of the magnetic field for inducing the re-entrant superconducting phases [13], which indicates that the STM measurements in this work are within a single superconducting phase of $UTe_2$ (SC1 phase). The SC1 phase in $UTe_2$ emerges upon cooling from a nearly ferromagnetic state with crystallographic *b* axis as the magnetic hard axis [11, 42]. When an out-of-plane external magnetic field is applied, the magnetic fields enter in the form of vortices consisting of magnetic fluxes and circulating supercurrent. The STM measures the vortex-core states which are distributed on the length scale of coherence length (ξ). The magnetic fields decay from the vortex core on the length scale of London penetration depth (λ) [Figs. 4(c) and 4(e)] [23]. For the U-based superconductors, the London penetration depth can be several hundreds of nanometers or even larger [43].

Around the vortex core region, the magnetic moments from the U *f*-electrons can be polarized by the magnetic fields within the vortex, which induces magnetization (**M**). Since the *b* axis is the magnetic hard axis of $UTe_2$, when the magnetic field is applied perpendicular to the (0−11) surface, the magnetization would rotate toward the *c* axis from the magnetic field direction [Figs. 4(c) and 4(e)]. This would induce surface bound current (**K**) which is calculated by $\mathbf{M} \times \mathbf{n}$ (where **n** is the surface normal vector). In this situation, on the left side of the vortex core, the direction of the bound current is along the same direction of the supercurrent [Fig. 4(d)]. However, on the right side of the vortex, their flow directions are opposite [Fig. 4(d)]. This phenomenon would result in smaller net supercurrent on the right side of the vortex core than on the left side. The following question is: how does this induce the asymmetric superconducting quasiparticle spectrum?

Due to the Doppler shift, current in superconductors weakens pair correlations and shifts the energy of Bogoliubov quasiparticle excitations [44, 45]. In our case, one possible scenario is that the superfluid velocity on the right side of the vortex core ($\mathbf{v_R} = \mathbf{v_s} - \mathbf{v_K}$) is smaller than that on the left side ($\mathbf{v_L} = \mathbf{v_s} + \mathbf{v_K}$), where $\mathbf{v_s}$ and $\mathbf{v_K}$ are the superfluid velocities related to the supercurrent and surface bound current, respectively. Therefore, the Doppler shift effect on the right side of the vortex core is weaker than on the left side, which gives rise to the asymmetric d*I*/d*V* spectra. When reversing the direction of the out-of-plane magnetic fields, the magnetization direction within the vortex also changes [Figs. 4(e) and 4(f)]. In this case, both the direction of the bound current and the circulation direction of the supercurrent reverse, which keeps the asymmetry unchanged. This is also consistent with the magnetic-field-direction dependent d*I*/d*V* maps shown in Figs. 4(a) and 4(b).

Coexistence of magnetization and the superconducting gap near the vortex core supports the spin-triplet superconductivity in $UTe_2$. General symmetry analysis for odd-parity spin-triplet superconducting gap function of $UTe_2$ can be divided into two classes: chiral and nonchiral [46]. So far, the experimental identification of the superconducting gap symmetry of $UTe_2$ remains still under debate [17, 18, 21, 47, 48]. It can be regarded that the spatial configuration around a superconducting vortex, separating the superconducting bulk from the non-superconducting vortex core, is topologically equivalent to a superconductor with left and right boundaries. The symmetry breaking observed here is analogous to the previously reported chiral boundary states at the step edges of $UTe_2$ [17], and our proposed explanation for the magnetization-induced asymmetry around the vortex core indicates that the magnetic properties at the step edges of $UTe_2$ may play an important role in the chiral boundary modes in $UTe_2$.

**CONCLUSION**

Our STM data reveal many intriguing features for the vortex-core states and vortex lattice in $UTe_2$. The elongation behavior of the robust zero-energy vortex-core state should be a combined effect of the Fermi surface anisotropy and the superconducting gap structure of $UTe_2$. However, whether this zero-energy vortex-core state is the Majorana zero mode in a spin-triplet superconductor needs further experimental investigations. We detect the asymmetric d*I*/d*V* spectra on the two sides of the elongated vortex core, which is induced by the magnetization-induced bound current. Our findings also provide a new clue for understanding the chiral boundary modes in $UTe_2$. The enhanced depth of superconducting gap and coherence peak on one side of the vortex core is extremely special, and further theoretical modelling is needed to reveal its origins and implications. We also expect that this kind of symmetry breaking in the vortex should be a general phenomenon for superconductors with strong and anisotropic paramagnetism.

*Note added.* After submitting the manuscript, we became aware of the other two STM studies about the superconducting vortices in $UTe_2$ [49, 50].

**METHODS**

**Single-crystal growth** High-quality $UTe_2$ crystals were successfully synthesized using the molten salt flux (MSF) method [51]. Prior to the preparation, natural uranium metal (3N) was polished with an electric sander to eliminate surface oxides, and then cleaned with alcohol and acetone. High-purity NaCl (4N) and KCl (4N) were finely ground, and then baked in an oven at 120 °C for 24 hours to remove moisture. Initially, precise amounts of high-purity uranium, tellurium powder (5N), NaCl and KCl were weighed and mixed in a molar ratio of 1 : 1.71 : 20 : 20. This mixture was placed into a small $Al_2O_3$ crucible and subsequently loaded into a 13-cm-long quartz ampoule with an inner diameter of 20 mm, which was then sealed under a pressure of approximately $10^{-3}$ Pa. The sealed ampoule was introduced into a shaft furnace, where the temperature was gradually increased to 950 °C over 24 hours and maintained for another 24 hours. Afterwards, the temperature was adjusted to 650 °C over a period of two to three weeks to promote crystal growth. After cooling to room temperature, the content in the $Al_2O_3$ crucible was soaked in deionized water for 24 hours to remove NaCl and KCl, resulting in the successful production of the rod-like $UTe_2$ single crystals (Supplemental Material Fig. S1).

**STM/STS measurements** STM experiments were conducted with an ultralow-temperature STM system at the base temperature of 30 mK and with the effective electronic temperature of ~200 mK (Unisoku 1600) [52]. STM measurements were performed with the chemically etched tungsten tips. The tungsten tips were flashed by electron-beam bombardment for two minutes before use. The $UTe_2$ single crystal sample was cleaved at 77 K, and then transferred immediately into the STM head for measurement. The d$I$/d$V$ spectra were acquired using a standard lock-in technique at a modulation frequency of 910 Hz.

## SUPPLEMENTARY DATA

Supplementary data are available at *NSR* online.

## ACKNOWLEDGEMENTS

We thank Vidya Madhavan, Ziqiang Wang, Kazushige Machida and Yin Zhong for fruitful discussions.

## FUNDING

S.Y. received the financial support from the National Key Research and Development Program of China (2022YFA1402703 and 2020YFA0309602) and the start-up funding from ShanghaiTech University. C.Z. received the financial support from Outstanding Young Researcher Scheme of Hunan Province (2023JJ10051). The work at University of Oxford was supported by the Synergetic Extreme Condition User Facility (SECUF, https://cstr.cn/31123.02.SECUF).

## AUTHOR CONTRIBUTIONS

S.Y. and Y.C. conceived the experiments. Z.Y. and F.Z. obtained the STM data. X.C. developed the magnetization model. B.Z., N.L. and C.Z. provided the $UTe_2$ single crystals. Z.Y., F.Z. and S.Y. performed the data analysis and wrote the paper with the input from all authors.

***Conflict of interest statement.*** None declared.

## REFERENCES

1. Mackenzie AP, Maeno Y. The superconductivity of $Sr_2RuO_4$ and the physics of spin-triplet pairing. *Reviews of Modern Physics*. 2003; **75**(2): 657-712.
2. Read N, Green D. Paired states of fermions in two dimensions with breaking of parity and time-reversal symmetries and the fractional quantum Hall effect. *Physical Review B*. 2000; **61**(15): 10267-10297.
3. Vakaryuk V, Leggett AJ. Spin Polarization of Half-Quantum Vortex in Systems with Equal Spin Pairing. *Physical Review Letters*. 2009; **103**(5): 057003.
4. Hsieh TH, Fu L. Majorana Fermions and Exotic Surface Andreev Bound States in Topological Superconductors: Application to $Cu_xB_2Se_3$. *Physical Review Letters*. 2012; **108**(10): 107005.
5. Salomaa MM, Volovik GE. Quantized vortices in superfluid $^3$He. *Reviews of Modern Physics*. 1987;

**59**(3): 533-613.
6. Tsutsumi Y, Machida K. Topological spin texture and *d*-vector rotation in spin-triplet superconductors: A case of $UTe_2$. *Physical Review B*. 2024; **110**(6): L060507.
7. Stewart GR, Fisk Z, Willis JO *et al.* Possibility of Coexistence of Bulk Superconductivity and Spin Fluctuations in $UPt_3$. *Physical Review Letters*. 1984; **52**(8): 679-682.
8. Saxena SS, Agarwal P, Ahilan K *et al.* Superconductivity on the border of itinerant-electron ferromagnetism in $UGe_2$. *Nature*. 2000; **406**(6796): 587-592.
9. Aoki D, Huxley A, Ressouche E *et al.* Coexistence of superconductivity and ferromagnetism in URhGe. *Nature*. 2001; **413**(6856): 613-616.
10. Huy NT, Gasparini A, de Nijs DE *et al.* Superconductivity on the Border of Weak Itinerant Ferromagnetism in UCoGe. *Physical Review Letters*. 2007; **99**(6): 067006.
11. Ran S, Eckberg C, Ding Q-P *et al.* Nearly ferromagnetic spin-triplet superconductivity. *Science*. 2019; **365**(6454): 684-687.
12. Aoki D, Nakamura A, Honda F *et al.* Unconventional Superconductivity in Heavy Fermion $UTe_2$. *Journal of the Physical Society of Japan*. 2019; **88**(4): 043702.
13. Ran S, Liu IL, Eo YS *et al.* Extreme magnetic field-boosted superconductivity. *Nature Physics*. 2019; **15**(12): 1250-1254.
14. Aoki D, Brison J-P, Flouquet J *et al.* Unconventional superconductivity in $UTe_2$. *Journal of Physics: Condensed Matter*. 2022; **34**(24): 243002.
15. Nakamine G, Kinjo K, Kitagawa S *et al.* Anisotropic response of spin susceptibility in the superconducting state of $UTe_2$ probed with $^{125}$Te-NMR measurement. *Physical Review B*. 2021; **103**(10): L100503.
16. Sundar S, Gheidi S, Akintola K *et al.* Coexistence of ferromagnetic fluctuations and superconductivity in the actinide superconductor $UTe_2$. *Physical Review B*. 2019; **100**(14): 140502(R).
17. Jiao L, Howard S, Ran S *et al.* Chiral superconductivity in heavy-fermion metal $UTe_2$. *Nature*. 2020; **579**(7800): 523-527.
18. Hayes IM, Wei DS, Metz T *et al.* Multicomponent superconducting order parameter in $UTe_2$. *Science*. 2021; **373**(6556): 797-801.
19. Bae S, Kim H, Eo YS *et al.* Anomalous normal fluid response in a chiral superconductor $UTe_2$. *Nature Communications*. 2021; **12**(1): 2644.
20. Ishihara K, Roppongi M, Kobayashi M *et al.* Chiral superconductivity in $UTe_2$ probed by anisotropic low-energy excitations. *Nature Communications*. 2023; **14**(1).
21. Matsumura H, Fujibayashi H, Kinjo K *et al.* Large Reduction in the a-axis Knight Shift on $UTe_2$ with Tc = 2.1 K. *Journal of the Physical Society of Japan*. 2023; **92**(6): 063701.
22. Lewin SK, Frank CE, Ran S *et al.* A review of $UTe_2$ at high magnetic fields. *Reports on Progress in Physics*. 2023; **86**: 114501.
23. Abrikosov AA. The magnetic properties of superconducting alloys. *Journal of Physics and Chemistry of Solids*. 1957; **2**(3): 199-208.
24. Hess HF, Robinson RB, Dynes RC *et al.* Scanning-Tunneling-Microscope Observation of the Abrikosov Flux Lattice and the Density of States near and inside a Fluxoid. *Physical Review Letters*. 1989; **62**(2): 214-216.
25. Hess HF, Robinson RB, Waszczak JV. Vortex-core structure observed with a scanning tunneling microscope. *Physical Review Letters*. 1990; **64**(22): 2711-2714.
26. Song C-L, Wang Y-L, Cheng P *et al.* Direct Observation of Nodes and Twofold Symmetry in FeSe

Superconductor. *Science*. 2011; **332**(6036): 1410-1413.
27. Suderow H, Guillamón I, Rodrigo JG *et al.* Imaging superconducting vortex cores and lattices with a scanning tunneling microscope. *Superconductor Science and Technology*. 2014; **27**(6): 063001.
28. Wang D, Kong L, Fan P *et al.* Evidence for Majorana bound states in an iron-based superconductor. *Science*. 2018; **362**(6412): 333-335.
29. Chen M, Chen X, Yang H *et al.* Discrete energy levels of Caroli-de Gennes-Matricon states in quantum limit in $FeTe_{0.55}Se_{0.45}$. *Nature Communications*. 2018; **9**(1): 970.
30. Aishwarya A, May-Mann J, Raghavan A *et al.* Magnetic-field-sensitive charge density waves in the superconductor $UTe_2$. *Nature*. 2023; **618**(7967): 928-933.
31. Gu Q, Carroll JP, Wang S *et al.* Detection of a pair density wave state in $UTe_2$. *Nature*. 2023; **618**(7967): 921-927.
32. Aishwarya A, May-Mann J, Almoalem A *et al.* Melting of the charge density wave by generation of pairs of topological defects in $UTe_2$. *Nature Physics*. 2024; **20**(6): 964-969.
33. LaFleur A, Li H, Frank CE *et al.* Inhomogeneous high temperature melting and decoupling of charge density waves in spin-triplet superconductor $UTe_2$. *Nature Communications*. 2024; **15**(1): 4456.
34. Ajeesh MO, Bordelon M, Girod C *et al.* Fate of Time-Reversal Symmetry Breaking in $UTe_2$. *Physical Review X*. 2023; **13**(4): 041019.
35. Wei DS, Saykin D, Miller OY *et al.* Interplay between magnetism and superconductivity in $UTe_2$. *Physical Review B*. 2022; **105**(2): 024521.
36. Rosa PFS, Weiland A, Fender SS *et al.* Single thermodynamic transition at 2 K in superconducting $UTe_2$ single crystals. *Communications Materials*. 2022; **3**(1): 33.
37. Thomas SM, Stevens C, Santos FB *et al.* Spatially inhomogeneous superconductivity in $UTe_2$. *Physical Review B*. 2021; **104**(22): 224501.
38. Azari N, Yakovlev M, Rye N *et al.* Absence of Spontaneous Magnetic Fields due to Time-Reversal Symmetry Breaking in Bulk Superconducting $UTe_2$. *Physical Review Letters*. 2023; **131**(22): 226504.
39. Aoki D, Sheikin I, Marquardt N *et al.* High Field Superconducting Phases of Ultra Clean Single Crystal $UTe_2$. *Journal of the Physical Society of Japan*. 2024; **93**(12): 123702.
40. Li M, Li G, Cao L *et al.* Ordered and tunable Majorana-zero-mode lattice in naturally strained LiFeAs. *Nature*. 2022; **606**(7916): 890-895.
41. Tokiwa Y, Sakai H, Kambe S *et al.* Anomalous vortex dynamics in the spin-triplet superconductor $UTe_2$. *Physical Review B*. 2023; **108**(14): 144502.
42. Rosuel A, Marcenat C, Knebel G *et al.* Field-Induced Tuning of the Pairing State in a Superconductor. *Physical Review X*. 2023; **13**(1): 011022.
43. Gross F, Andres K, Chandrasekhar BS. Experimental determination of the absolute value of the London penetration depth in the heavy fermion superconductors $UBe_{13}$ und $UPt_3$. *Physica C: Superconductivity and its Applications*. 1989; **162-164**: 419-420.
44. Zhu Z, Papaj M, Nie X-A *et al.* Discovery of segmented Fermi surface induced by Cooper pair momentum. *Science*. 2021; **374**(6573): 1381-1385.
45. Liu X, Chong YX, Sharma R *et al.* Atomic-scale visualization of electronic fluid flow. *Nature Materials*. 2021; **20**(11): 1480-1484.
46. Kallin C, Berlinsky J. Chiral superconductors. *Reports on Progress in Physics*. 2016; **79**(5): 054502.
47. Iguchi Y, Man H, Thomas SM *et al.* Microscopic Imaging Homogeneous and Single Phase Superfluid Density in $UTe_2$. *Physical Review Letters*. 2023; **130**(19): 196003.
48. Suetsugu S, Shimomura M, Kamimura M *et al.* Fully gapped pairing state in spin-triplet

superconductor $UTe_2$. *Science Advances*. 2024; **10**(6): eadk3772.
49. Sharma N, Toole M, McKenzie J *et al.* Observation of Persistent Zero Modes and Superconducting Vortex Doublets in $UTe_2$. 2025: arXiv:2503.17450.
50. Yin R, Li Y, Du Z *et al.* Yin-Yang vortex on $UTe_2$ (011) surface. 2025: arXiv:2503.21506.
51. Sakai H, Opletal P, Tokiwa Y *et al.* Single crystal growth of superconducting $UTe_2$ by molten salt flux method. *Physical Review Materials*. 2022; **6**(7): 073401.
52. Zhong R, Yang Z, Wang Q *et al.* Spatially Dependent in-Gap States Induced by Andreev Tunneling through a Single Electronic State. *Nano Letters*. 2024; **24**(28): 8580-8586.

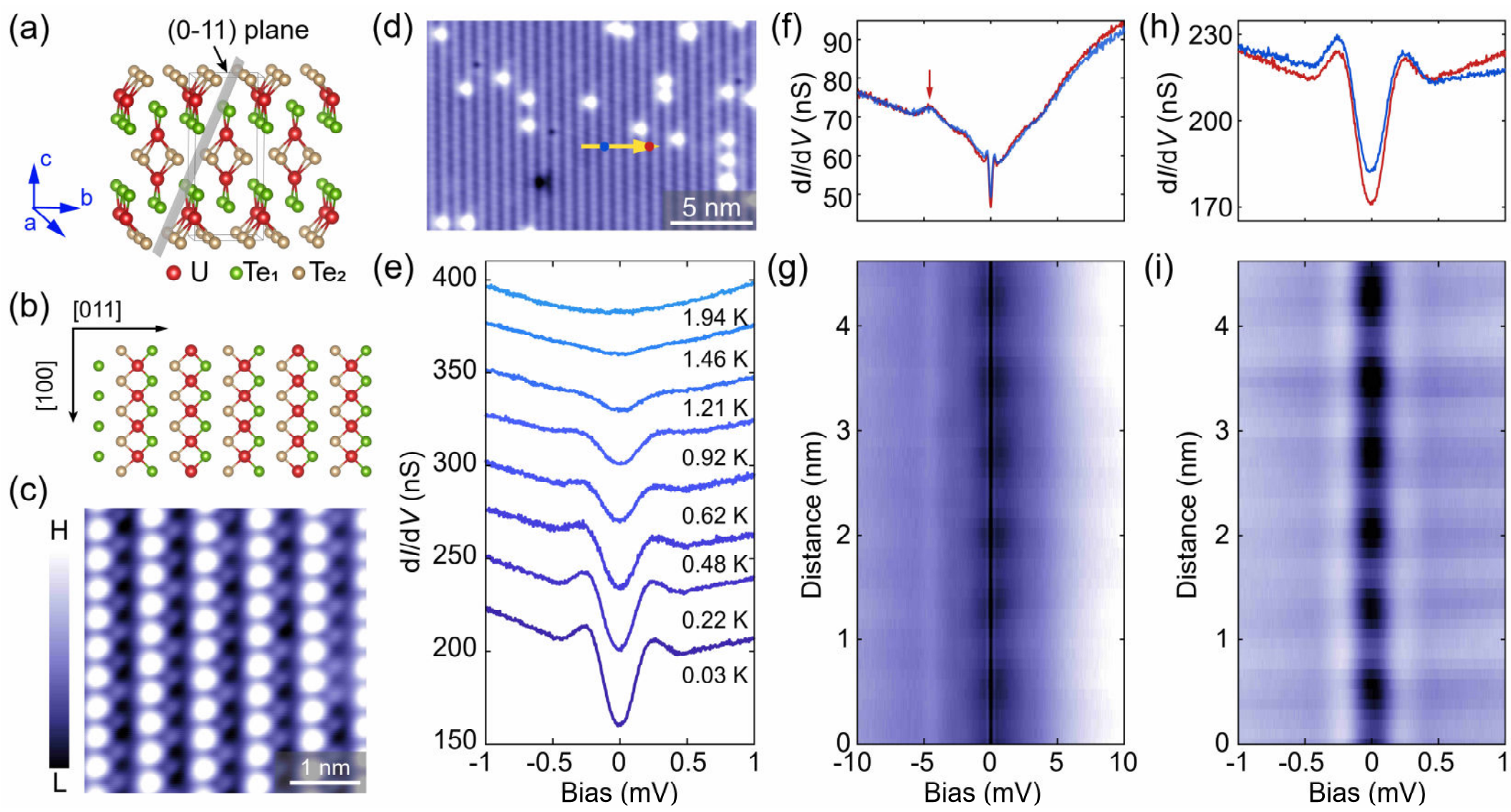

**Figure 1.** Superconductivity in $UTe_2$. (a) Crystal structure of $UTe_2$ with the cleavage plane shown by the grey rectangle. (b) Schematic for the structure of the (0−11) plane, which shows the $Te_1$ and $Te_2$ rows with the underlying U atoms. (c) High-resolution STM topography on the (0−11) surface where the $Te_1$ and $Te_2$ rows appear as alternating bright and dark atomic chains. (d) Typical STM topography on the (0−11) surface of $UTe_2$. (e) Variable-temperature $dI/dV$ spectra on the (0−11) surface, showing the evolution of the superconducting gap with temperature. (f),(g) $dI/dV$ linecut profile taken along the yellow arrow in (d) with ±10 mV energy range (g) (Setpoint: $V_s = -10$ mV, $I = 700$ pA). $dI/dV$ spectra in (f) are taken on the bright Te chain (blue) and between the bright Te chains (red), as marked by the blue and red dots in (d). (h),(i) Similar to (f) and (g), the $dI/dV$ spectra are taken with ±1 mV energy range (Setpoint: $V_s = -3$ mV, $I = 700$ pA).

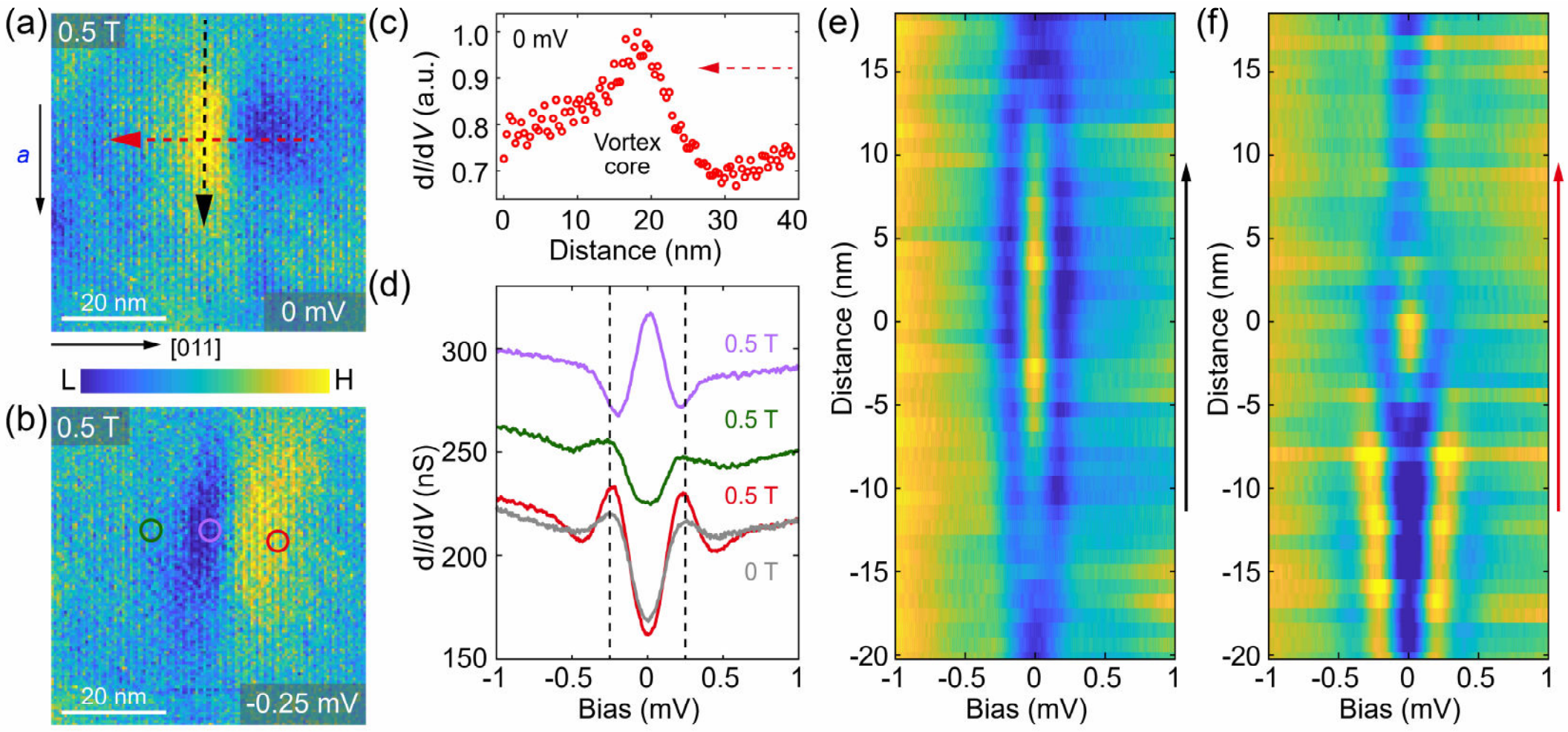

**Figure 2.** Observation of vortex and vortex-core state. (a) Zero-energy d$I$/d$V$ map taken in a magnetic field $B$ = 0.5 T. (b) d$I$/d$V$ map taken in the same area as in (a) with 0.5 T and −0.25 mV. (c) Zero-energy d$I$/d$V$ signal along the red dashed arrow in (a). (d) d$I$/d$V$ spectra taken at the center of the vortex (purple), on the left (green) and right (red) sides of the vortex core with 0.5 T, and the positions for these spectra are marked by the colored circles in (b). The spectra are vertically offset for clarity. The grey d$I$/d$V$ spectrum taken with zero magnetic field is shown for comparison. (e),(f) d$I$/d$V$ linecut profiles taken along the black (e) and red (f) dashed arrows in (a), which shows the evolution of the zero-energy vortex-core state inside the superconducting gap. The d$I$/d$V$ maps and d$I$/d$V$ spectra in this figure are taken with setpoint $V_s$ = −3 mV and $I$ = 700 pA.

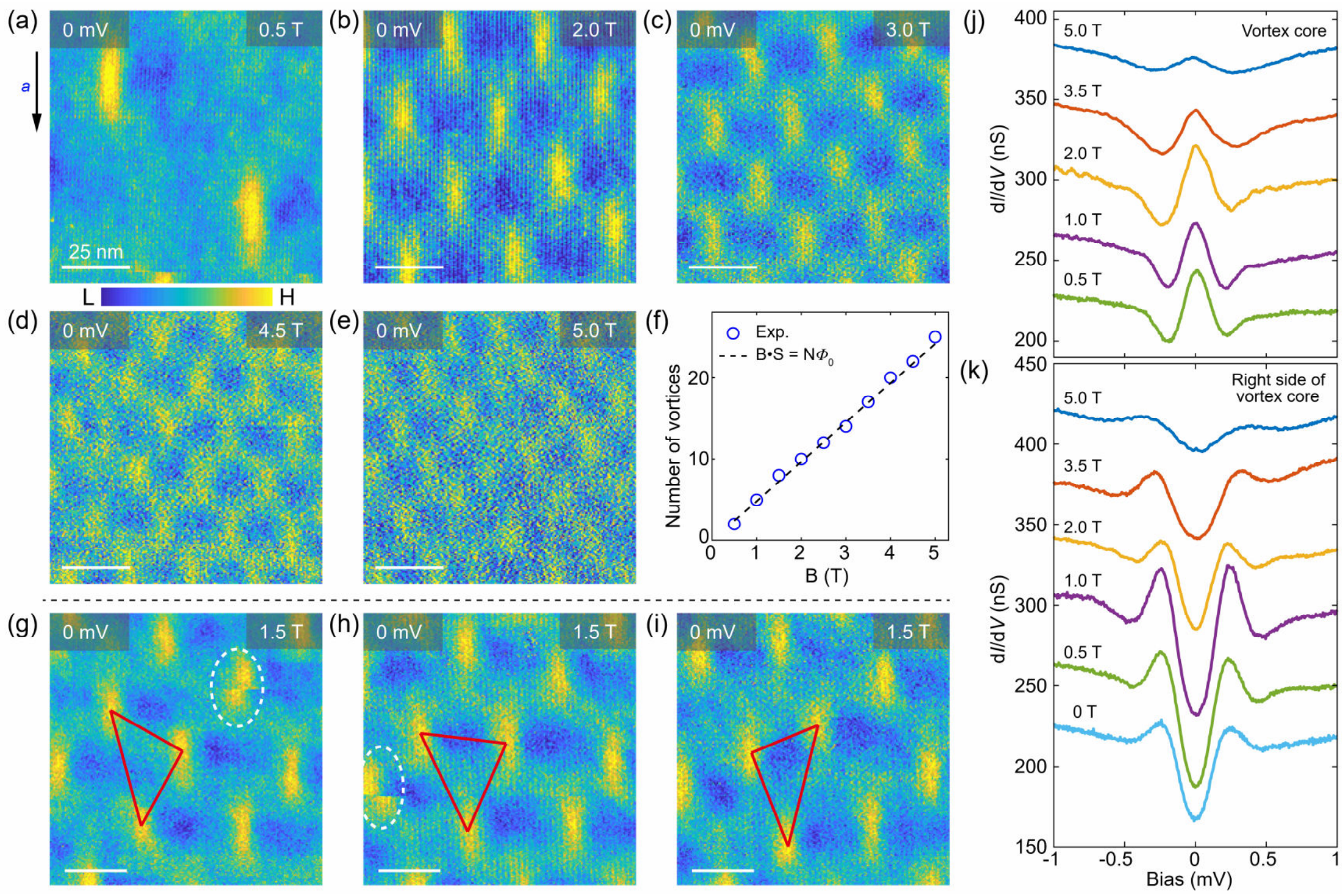

**Figure 3.** Evolution of vortex lattice as a function of external magnetic field. (a)-(e) Zero-energy d$I$/d$V$ maps taken at 0.5 T (a), 2.0 T (b), 3.0 T (c), 4.5 T (d), and 5.0 T (e) magnetic fields. (f) Number of vortices (blue circles) as a function of magnetic fields perpendicular to the surface. The dashed line represents the theoretical values of the number of vortices in a 100 × 100 nm$^2$ area. (g)-(i) A series of zero-energy d$I$/d$V$ maps taken with 1.5 T magnetic fields on the same area within 48 hours. The red triangles indicate the relative positions of the vortex cores. The dashed ellipses mark the moved vortex cores during the d$I$/d$V$ map measurements. Scale bar in (a)-(i): 25 nm. (j) Magnetic-field dependence of the d$I$/d$V$ spectra taken at the center of the vortex. (k) Magnetic-field dependence of the d$I$/d$V$ spectra taken on the right side of the vortex and the d$I$/d$V$ spectrum taken with zero magnetic field. The spectra in (j) and (k) are vertically offset for clarity. The d$I$/d$V$ maps and d$I$/d$V$ spectra in this figure are taken with setpoint $V_s = -3$ mV and $I = 700$ pA.

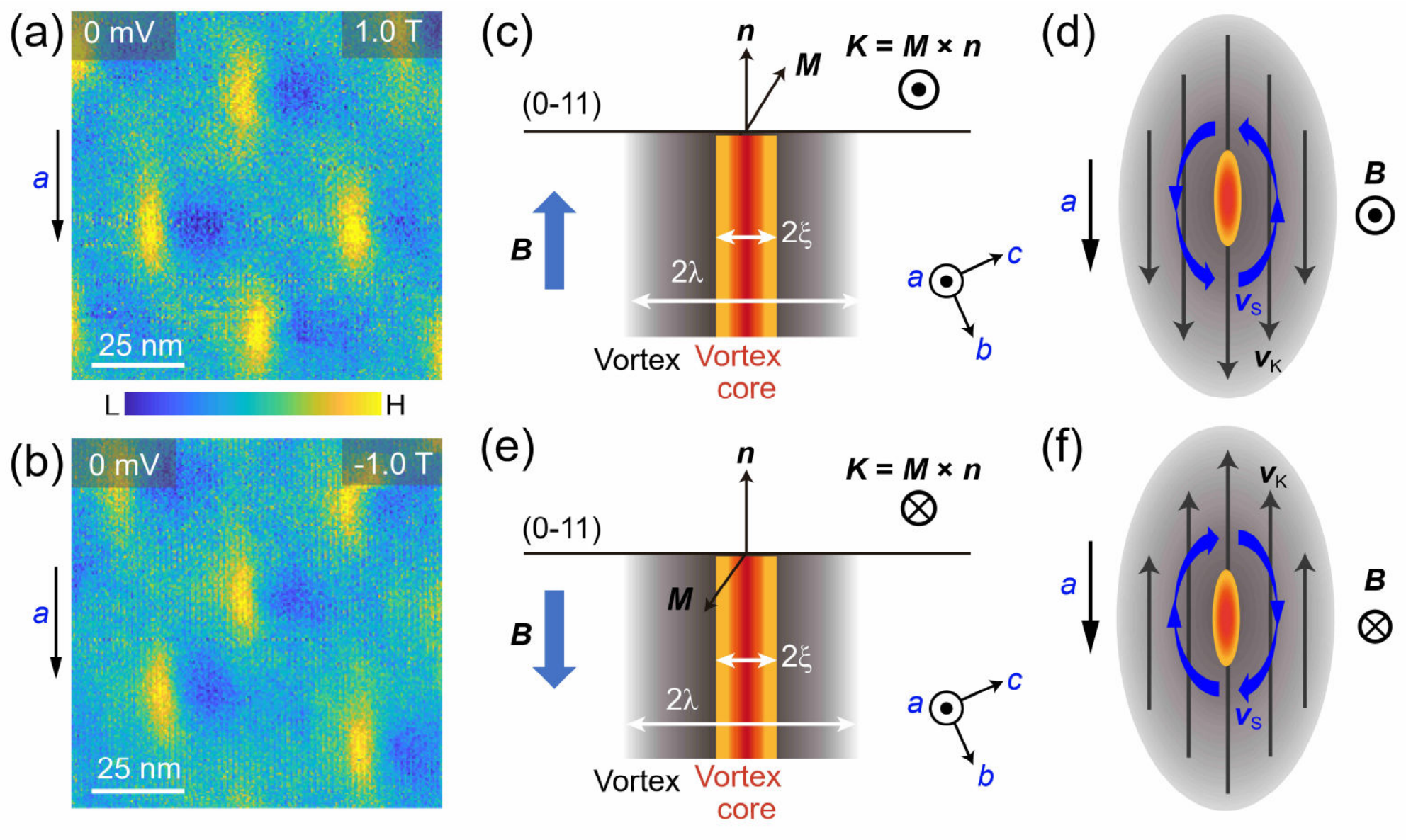

**Figure 4.** Magnetic-field-direction independence and phenomenology of the symmetry breaking near the vortex core. (a),(b) Zero-energy d$I$/d$V$ maps taken in the same area with 1 T (a) and −1 T (b) magnetic fields (Setpoint: $V_s = -3$ mV, $I = 700$ pA). (c) Schematic showing the magnetization (**M**) with magnetic field along the surface normal vector (**n**) and the bound current (**K** = **M** × **n**). ξ denotes the coherence length, and λ is the London penetration depth in $UTe_2$. (d) Schematic showing the bound current (**K** = **M** × **n**) with superfluid velocity ($\mathbf{v}_K$) and the circulating supercurrent with superfluid velocity ($\mathbf{v}_s$) around the vortex core. The red-orange and grey ovals represent the vortex core and the vortex in $UTe_2$, respectively. (e),(f) Similar to (c) and (d), but with the magnetic field along the opposite direction.